\documentclass[12pt,dviwindows]{article}

\usepackage{color,epsf,graphicx}
\usepackage{amsmath,amssymb}
\usepackage{psfrag}
\usepackage[colorlinks=true,citecolor=blue,linkcolor=red]{hyperref}

\newcounter{comment}

{\refstepcounter{comment}%
\begin{quote}
\ttfamily\small$\blacksquare$ \textbf{\underline{Comment} $\sharp$\thecomment:}}%
{\end{quote}}

{
\begin{quote}
\ttfamily\small$\blacktriangleright$ \textbf{\underline{Reply} $\sharp$\thecomment:}}%
{\end{quote}}

\newcommand{\insertfig}[2]{\mbox{\epsfxsize=#1cm \epsfbox{#2.eps}}}

\font\cmss=cmss12 
\def\1{\hbox{{1}\kern-.25em\hbox{l}}}
\def\bfZ{\relax{\hbox{\cmss Z\kern-.4em Z}}}

\def\ru1{\rule[-0.4truecm]{0mm}{1truecm}}

\begin{document}

\begin{titlepage}

\centerline{\large \bf  Implication of the overlap representation
for modelling} \centerline{\large \bf  generalized parton
distributions}

\vspace{7mm}

\centerline{\bf D.~S.~Hwang$^{a}$ and  D.~M\"uller$^{b}$}

\vspace{7mm}

\vspace{4mm} \centerline{\it $^a$Department of Physics, Sejong
University } \centerline{\it Seoul 143--747, South Korea}

\vspace{4mm} \centerline{\it $^b$Institut f\"ur Theoretische
Physik II, Ruhr-Universit\"at Bochum} \centerline{\it D-44780
Bochum, Germany}

\vspace{20mm}

\centerline{\bf Abstract}

\vspace{10mm}

\noindent Based on a field theoretically inspired model of
light-cone wave functions, we derive valence-like generalized
parton distributions and their double distributions  from the wave
function overlap in the parton number conserved $s$-channel. The
parton number changing contributions in the $t$-channel are
restored from duality. In our construction constraints of
positivity and polynomiality are simultaneously satisfied and it
also implies a model dependent relation between generalized parton
distributions and transverse momentum dependent parton
distribution functions. The model predicts that the $t$-behavior
of resulting hadronic amplitudes depends on the Bjorken
variable $x_{\rm Bj}$. We also propose an improved ansatz for
double distributions  that embeds this property.

\vspace{0.5cm}

\noindent \vspace*{12mm}


\vspace{3mm}


\end{titlepage}

\noindent i.~Since the  discovery that the proton is a composed
system, enormous amount of efforts have been made in order to
reveal its contents and understand the dynamics of its
constituents. Various frameworks and models have been
simultaneously proposed; often based on the quantum mechanical
concept of the proton wave function. Factorization theorems,
derived in the framework of perturbative Quantum Chromodynamics
(QCD), are the basis to relate non-perturbative parton
distribution functions (PDFs), distribution amplitudes, and
generalized parton distributions (GPDs) with experimental
observables. It is obvious that these non-perturbative quantities
are somehow related to the proton wave function. However,
quantifying this relation starts with the conceptional difficulty
of defining the bound state problem in the relativistic quantum
field theory, and it ends up with the difficulty in matching the
evaluated (generalized) parton distributions with those that are
defined implicitly in the perturbative factorization approach.

The idea to write down the proton wave function in terms of the
partonic degrees of freedom was spelled out in the early days
\cite{DreLevYan69,DreYan69,Wes70}. Thinking of the proton as a
bunch of partons that move nearly on the light-cone, e.g.,
specified by $n^\mu=(1,0,0,-1)$, allows more easily to establish
the desired link, see Refs.~\cite{BroDre80,BroPauPin97} and
references therein. This leads to the concept of light-cone (LC)
wave functions $\psi^{\uparrow,\downarrow}_n(X_i,{\bf k}_{\perp
i},\lambda_i)$. They are the probability amplitudes for their
corresponding $n$-parton states $|n, p^+_i, {\bf p}_{\perp
i},\lambda_i\rangle$, which build up the proton state with LC
helicity $S=\{+1/2 (\uparrow),-1/2(\downarrow)\}$:
\begin{eqnarray}
\label{Def-ProSta} |P,S=\{\uparrow,\downarrow\}\rangle = \sum_n
\int\![dX\, d^2{\bf k}]_n\, \psi^{\uparrow,\downarrow}_n(X_i,{\bf
k}_\perp,\lambda_i)\, \prod_{j=1}^n \frac{1}{\sqrt{X_j}}\, 
|n, X_i P^+, X_i {\bf P}_\perp + {\bf
k}_{\perp i},\lambda_i\rangle\,,
\end{eqnarray}
where we used a shorthand  for the $n$-parton phase space:
\begin{equation}
 [dx\, d^2{\bf k}_{\perp}]_n=\prod_{i=1}^n 
\frac{dX_i d^2 {\bf k}_{\perp i}}{16\pi^3}
16\pi^3 \delta\!\left(1-\sum_{j=1}^n X_i\right)
\delta^{(2)}\!\left(\sum_{j=1}^n {\bf k}_{\perp i}\right)\,.
\end{equation}
The LC wave functions depend on the longitudinal momentum
fractions $X_i = k^+_i/P^+$ (the plus component of a four-vector
$V^\mu$ is $V^+=V^0+V^{3} = n\cdot V$), the transverse momenta
${\bf k}_{\perp i}$, and the LC helicities $\lambda_i$. They are
determined from the eigenvalue problem for the LC Hamiltonian
$P^-$:
\begin{eqnarray}
P^- |P,S\rangle  = \frac{M^2}{P^+} |P,S\rangle \,,  \quad
\mbox{with}\quad P^-=P^0-P^3\,,\; P^+=P^0+P^3\,,\;  {\bf
P}_\perp=0\,,
\end{eqnarray}
which can be derived from the QCD Lagrangian. However, in practice
the QCD dynamics is not well understood and it remains very
challenging to develop this concept to a stage at which it can be
used for quantitative evaluations of physical observables or
parton distributions \cite{BroPauPin97}. It is common to pin down
LC wave functions by comparing their resulting model predictions
with experimental observations. Certainly, so far this concept is
extensively elaborated, it connects different non-perturbative
quantities, like PDFs, transverse momentum dependent parton
distribution functions (TMDs), distribution amplitudes, form
factors, and GPDs to a more universal object.

Our study is mainly devoted to GPDs, which are accessible from
hard-exclusive reactions, e.g., electroproduction of mesons and
photon. They arise from the non-diagonal overlap of LC wave
functions, and therefore contain a maximum of information about
the proton wave function, compared to other non-perturbative
quantities. For instance, they are reducible to elastic form
factors and PDFs. Field theoretically they are defined as
off-diagonal matrix elements of two field operators that live on
the light cone \cite{MueRobGeyDitHor94,Rad96,Ji96a}. Describing
the initial and final proton states, with given momenta $P_1$ and
$P_2=P_1-\Delta$ and LC helicities $S_1$ and $S_2$, respectively,
in terms of the LC wave functions (\ref{Def-ProSta}), one can
straightforwardly derive wave function overlap representations of
GPDs for the partonic $s$-channel exchange
\cite{DieFelJakKro98,BroDieHwa00,DieFelJakKro00}.  In this
partonic process the number of partons is conserved and the
momentum fraction $x$ of the struck quark is larger than the
skewness parameter $\eta = (P_1^+-P_2^+)/(P_1^++P_2^+) >0$ (up to
a minus sign we use the variable conventions of
\cite{MueRobGeyDitHor94}). In this {\em outer} region of $x$, the
GPDs $H$ and $E$ (Ji's conventions \cite{Ji98}) read
\begin{eqnarray}
\label{OveRep-DGLAP-H} &&\hspace{-1cm} \left(H -
\frac{\eta^2}{1-\eta^2} E\right)(x\ge \eta,\eta,t) =
\\
&&\hspace{-0cm} \frac{2-\zeta}{2\sqrt{1-\zeta}} \sum_n
\sum_{\lambda_i} \sqrt{1-\zeta}^{2-n} \int\! [dX\, d^2{\bf
k}_{\perp}]_n\, \, \delta(X-X_1) \psi^{\uparrow
\ast}_n(X_i^\prime,{\bf k}^\prime_{\perp i},\lambda_i)
\psi^{\uparrow}_n(X_i,{\bf k}_{\perp i},\lambda_i)\,,
\nonumber\\
\label{OveRep-DGLAP-E} &&\hspace{-1cm}\frac{\Delta_1- i
\Delta_2}{2 M} E(x\ge \eta,\eta,t) =
\\
&&\hspace{-0cm} \sqrt{1-\zeta} \sum_{n} \sum_{\lambda_i}
\sqrt{1-\zeta}^{2-n} \int\! [dX\, d^2{\bf k}_{\perp}]_n\,
\, \delta(X-X_1) \psi^{\uparrow
\ast}_n(X_i^\prime,{\bf k}^\prime_{\perp i},\lambda_i)
\psi^{\downarrow}_n(X_i,{\bf k}_{\perp i},\lambda_i), \nonumber
\end{eqnarray}
where $\zeta =2\eta/(1+\eta)$, $X=(x+\eta)/(1+\eta)$, and the
momenta of the outgoing partons are
\begin{eqnarray}
&&  X_1^\prime= \frac{X_1-\zeta}{1-\zeta}\,, \quad {\bf
k}^\prime_{\perp 1} ={\bf k}_{\perp 1}- \frac{1-X_1}{1-\zeta} {\bf
\Delta}_\perp
\quad \mbox{for the struck quark},\\
&& X_i^\prime = \frac{X_i}{1-\zeta}\,, \quad\;\; \ {\bf
k}^\prime_{\perp i} ={\bf k}_{\perp i}+\frac{X_i}{1-\zeta} {\bf
\Delta}_\perp \quad\;\, \mbox{for the spectators $i=2,\dots,n$}.
\end{eqnarray}
Anti-quark GPDs are analogously defined with a negative momentum
fraction $x \le -\eta$. The {\em central} region, i.e., $ -\eta
\le x\le \eta$, arises from the $t$-channel process in which the
parton number changes from $n+2$ to $n$
\cite{BroDieHwa00,DieFelJakKro00}. Viewing this as a mesonic-like
$t$-channel exchange makes contact to Regge phenomenology
\cite{PolShu02,MueSch05,KumMuePas07}. Note that positivity
constraints, in its most general form \cite{Pob02}, should be  satisfied
in the overlap representations \cite{Pob03}, if they are not
spoiled by subtraction procedures. Indeed, this can be easily shown for
a two-body LC wave function, as used below.

Let us also remind of the constraints of Lorentz covariance for
(quark) GPD form factors. They are not invariant under general
Lorentz transformations, however,
they are built by a series of local twist-two operator matrix
elements, belonging to irreducible representations, that are
labelled by the spin $J \ge 1$. It turns out that Mellin moments
of GPDs $H+E$ and $E$ with the weight $x^{J-1}$ are polynomials in
$\eta$ of the order $J-1$ and $J$, respectively. Time reversal
invariance combined with hermiticity requires that these
polynomials are even \cite{Ji98}. These properties are manifestly
implemented in the double distribution (DD) representation of GPDs
\cite{MueRobGeyDitHor94,Rad96}:
\begin{eqnarray}
\label{Def-SpeRep} \left\{H+E\atop E \right\}(x,\eta,t) =
\int_{0}^1\!dy\; \int_{-1+y}^{1-y}\! dz\; \left\{ 1 \atop 1-x
\right\} \delta(x-y-\eta z )\; \left\{h+e \atop e \right\}(y,z,t),
\end{eqnarray}
where the DDs $e$ and $h$ are symmetric in $z$. The DD
representation  for $E$ is not uniquely defined
\cite{BelKirMueSch01,Ter01}; Eq.~(\ref{Def-SpeRep}) shows the one
which naturally occurs in our model studies, see below
Eq.~(\ref{eleeta}).

As explained above, the partonic interpretation of GPDs separates
the support in the central ($|x| \le \eta$) and outer ($  \eta \le
|x|$) regions. Residual Lorentz covariance, ensuring the
polynomiality of moments, requires that both regions are tied to
each other and that the functional form of GPD is constrained,
e.g., in the outer region the GPD $E$ is given by the integral
representation:
\begin{eqnarray}
\label{SpeRep-E-DGLAP} E(x \ge \eta,\eta,t) =
(1-x)\int_{\frac{x-\eta}{1-\eta}}^{\frac{x+\eta}{1+\eta}}\!\frac{dy}{\eta}\;
e(y,(x-y)/\eta,t)\,.
\end{eqnarray}
We consider both regions as dual to each other, i.e., knowing a
GPD in one region allows to restore it in the other. Hence, a GPD
can be entirely evaluated from the parton number conserved
$s$-channel overlap of LC wave functions. A constructive, however,
unwieldy method for the restoration of the central region is based
on Mellin moments and its inverse transformation \cite{MueSch05}.

In the following we utilize two-particle LC wave functions. They
serve us to describe the proton contents by a constituent quark
and scalar diquark, where the latter plays the role of an
collective spectator. Numerous investigations in this spirit, even
much more advanced ones with specific emphases, e.g.,
Refs.~\cite{Sch92,BolJakKroBerSte95,DieFelJakKro98,ChoJiKis01,TibMil01,
TibMil01a,BofPasTra02,AhmHonLiuTan06},
have been made in the past. The new aspect in our study is that we
take care of Lorentz constraints for the LC wave functions,
cf.~Refs.~\cite{MukMusPauRad02,JiMisRad06}. This allows us to
evaluate\footnote{Of course, this task is straightforwardly done
in a Lorentz covariant formalism \cite{TibDetMil04}.} DDs from the
parton number conserving overlap representations
(\ref{OveRep-DGLAP-H}) and (\ref{OveRep-DGLAP-E}). We outline the
generic features of such model, illuminate their $t$-dependence,
point out its restrictions, and overcome them by hand, yielding
improved DD and GPD ansatze.

\vspace{2mm}

\noindent ii.~The functional form of LC wave functions is dictated
by the underlying Lorentz symmetry, i.e., the longitudinal and
transversal variables are tied to each other in a certain but not
apparent manner.  Hence, writing down LC wave functions by hand
usually results in a violation of the GPD polynomiality property.
Note that this failure can not be fixed by taking into account the
particle number changing processes. The guidance for an
appropriate model comes from a perturbative calculation in lowest
order \cite{BroHwa98}. We employ the Yukawa theory  and have the
LC wave functions for four helicities,
\begin{eqnarray}
\label{Def-LC-WF1} &&\!\!\!\psi^{\uparrow}_{+1/2}(X,{\bf k}_\perp)
= \psi^{\downarrow}_{-1/2}(X,{\bf k}_\perp)
=\left(M+\frac{m}{X}\right) \varphi(x,{\bf k}_\perp)\,,
\\
\label{Def-LC-WF2} &&\!\!\!\psi^{\downarrow}_{+1/2}(X,{\bf
k}_\perp) = \frac{k^1- i k^2}{X} \varphi(X,{\bf k}_\perp) \,,\quad
\psi^{\uparrow}_{-1/2}(X,{\bf k}_\perp) = - \frac{k^1+ i k^2}{X}
\varphi(X,{\bf k}_\perp)\,,
\end{eqnarray}
in terms of a scalar function $\varphi(X,{\bf k}_\perp)$. This
scalar function arises from the spectator propagator in a triangle
Feynman diagram \cite{BroHwa98,BroHwaSch00} and so the underlying
Lorentz symmetry is respected. We generalize $\varphi(X,{\bf
k}_\perp)$ by an adjustment of its power behavior $p$:
\begin{eqnarray}
\label{Def-LC-WF3} \varphi(X,{\bf k}_\perp) = \frac{g
M^{2p}}{\sqrt{1-X}} X^{-p} \left( M^2 - \frac{{\bf
k}_\perp^2+m^2}{X} -\frac{{\bf k}_\perp^2+\lambda^2}{1-X}
\right)^{-p-1}\,,
\end{eqnarray}
where $M$, $\lambda$ and $m$ are the proton, spectator, and quark
masses, respectively. Note that the factor $X^{-p}$ takes care of
the proper Lorentz behavior and that the  Yukawa theory result is
for $p=0$.

The GPDs are now evaluated in the outer region from the overlap
representations (\ref{OveRep-DGLAP-H}) and (\ref{OveRep-DGLAP-E}),
\begin{eqnarray}
\label{Def-Hp} &&\hspace{-2cm} \left(H - \frac{\eta^2}{1-\eta^2}
E\right)( x\ge \eta,\eta,t) =
\\
&&  \frac{2-\zeta}{2\sqrt{1-\zeta}} \int\frac{d^2 {\bf
k}_\perp}{16 \pi^3} \left[\psi^{\uparrow\ast}_{+1/2}(X^\prime,{\bf
k}^\prime_\perp) \psi^{\uparrow}_{+1/2}(X,{\bf k}_\perp) +
\psi^{\uparrow\ast}_{-1/2}(X^\prime,{\bf k}^\prime_\perp)
\psi^{\uparrow}_{-1/2}(X,{\bf k}_\perp)\right],
\nonumber\\
\label{Def-Ep} &&\hspace{-2cm} \frac{\Delta^1-i \Delta^2}{2 M}
E(x\ge \eta,\eta,t)=
\\
&&
 \sqrt{1-\zeta} \int\frac{d^2 {\bf k}_\perp}{16 \pi^3}
\left[\psi^{\uparrow\ast}_{+1/2}(X^\prime,{\bf k}^\prime_\perp)
\psi^{\downarrow}_{+1/2}(X,{\bf k}_\perp) +
\psi^{\uparrow\ast}_{-1/2}(X^\prime,{\bf k}^\prime_\perp)
\psi^{\downarrow}_{-1/2}(X,{\bf k}_\perp)\right], \nonumber
\end{eqnarray}
for the two body wave functions
(\ref{Def-LC-WF1})--(\ref{Def-LC-WF3}). For $p>0$ the ${\bf
k}_\perp$ integrals are finite, while for $p=0$ the representation
(\ref{Def-Hp}) suffers from an ultraviolet divergence. We find for
the GPD $E$:
\begin{eqnarray}
\label{Res-E} E(x\ge\eta,\eta,t)= \frac{g^2}{(4\pi)^2}
\frac{2\Gamma(2p+1)}{\Gamma(p+1)^2} \int_{0}^1\!\!d\alpha
\frac{\left[(1-X)(1-X^\prime)\right]^{p+1} \left[\alpha
\overline{\alpha} \right]^p\left(\frac{m}{M} + X-\alpha
(1-X^\prime)\zeta\right) }{
\left[f\!\left(X|\overline{\alpha}\right)+f\!\left(X^\prime
|\alpha\right)+ \alpha \overline{\alpha} (1-X)(1-X^\prime)
\frac{t_{\rm min}-t}{M^2}\right]^{2 p+1} },
\end{eqnarray}
where $t_{\rm min}= -\zeta^2 M^2/(1-\zeta)$,
$\overline{\alpha}=1-\alpha$, and the mass terms are collected in
\begin{eqnarray}
f(X|\overline{\alpha})=\overline{\alpha} \left\{(1-X)
\frac{m^2}{M^2}+ X \frac{\lambda^2}{M^2} - X (1-X) \right\}.
\end{eqnarray}
The result for $H$ has a similar structure and will not be
displayed for shortness.

Since our model respects the underlying Lorentz symmetry, there
must be now a possibility to transform the overlap result
(\ref{Res-E}) into the form of  the DD representation
(\ref{SpeRep-E-DGLAP}). In fact, this can be simply achieved by a
linear variable transformation of the integration parameter
\begin{eqnarray}
\alpha = \frac{1}{2}\frac{1-\eta}{1-x} \left(1-y+\frac{x-y}{\eta
}\right)
\end{eqnarray}
and removing the residual skewness dependence by using $x= y +
\eta z$. We arrive at
\begin{eqnarray}
E(x\ge\eta,\eta,t)=
(1-x)\int_{\frac{x-\eta}{1-\eta}}^{\frac{x+\eta}{1+\eta}}\!\frac{dy}{\eta}\;
e\!\left(y, \frac{x-y}{\eta},t\right)\,, \label{egeeta}
\end{eqnarray}
where the DD is given by
\begin{eqnarray}
\label{Def-SpeFun-e} e\!\left(y,z,t\right) = N
\frac{\left(\frac{m}{M} + y\right) ((1-y)^2-z^2)^p}{\left[(1-y)
\frac{m^2}{M^2}+ y \frac{\lambda^2}{M^2} - y(1-y) -((1-y)^2-z^2)
\frac{t}{4 M^2}\right]^{2p+1}}\,.
\end{eqnarray}
Here we absorbed several factors, including  $g^2$,  in the
normalization constant $N$. From the DD (\ref{Def-SpeFun-e}) and
Eq.~(\ref{Def-SpeRep}), we  find $E$ for the central region,
arising from parton number changing processes:
\begin{eqnarray}
E(-\eta\le x\le\eta,\eta,t)=
(1-x)\int_{0}^{\frac{x+\eta}{1+\eta}}\! \frac{dy}{\eta}\;
e\!\left(y, \frac{x-y}{\eta},t\right)\,. \label{eleeta}
\end{eqnarray}

The evaluation of the GPD $H$ in terms of the DD $h$ goes along
the same line.  The combination $(H+E)(x,\eta,t)$ can be written
in the form of the DD representation (\ref{Def-SpeRep}) with
\begin{eqnarray}
(h+e)(y,z,t)  &\!\!\!=\!\!\! & N\frac{1-2 p}{4 p}
\frac{((1-y)^2-z^2)^p}{ \left[(1-y) \frac{m^2}{M^2}+ y
\frac{\lambda^2}{M^2} - y(1-y) -((1-y)^2-z^2) \frac{t}{4
M^2}\right]^{2p} }
\nonumber\\
&&  + N \frac{ \left[\frac{2 m}{M}+ y + y \frac{\lambda^2}{M^2} +
(2-y)  \frac{m^2}{M^2} \right]((1-y)^2-z^2)^p }{ 2\left[(1-y)
\frac{m^2}{M^2}+ y \frac{\lambda^2}{M^2} - y(1-y)
-((1-y)^2-z^2)\frac{t}{4 M^2}\right]^{2p+1}
}\phantom{\frac{\Bigg|}{\Big|}} \,. \label{Def-SpeFun-eh}
\end{eqnarray}
Finally, the overall constant $N$ is fixed by the normalization
condition:
\begin{eqnarray}
\label{Fix-Nor}
 \int_{-1}^1\!dx\; H(x,\eta,t=0)= \int_{0}^1\!dy\,\int_{-1+y}^{1-y}\!dz\, 
(h+y \, e)\!\left(y,z,t=0\right) =1\,.
\end{eqnarray}

\vspace{2mm}

\noindent iii.~We employ now our simple-minded model to evaluate
form factors, TMDs, PDFs, and GPDs as net contributions of
valence-like quarks in the proton, i.e., we deal with the isospin
combination
\begin{eqnarray}
 H(x,\eta,t) = H_{u_{\rm val}}(x,\eta,t) - H_{d_{\rm val}}(x,\eta,t)\,,\quad
 E(x,\eta,t) = E_{u_{\rm val}}(x,\eta,t) - E_{d_{\rm val}}(x,\eta,t)\,.
\end{eqnarray}
We have only three model parameters, i.e., power behavior $(p)$ of
the LC wave function, spectator ($\lambda$) and quark ($m$)
masses. Our goal in doing so is to explore this model and find its
restrictions.

We start with the electromagnetic form factors, which are
obtained  from the Drell-Yan-West formula or, equivalently,  by
integrating out the momentum fraction dependence in GPDs:
\begin{eqnarray}
\label{Def-FF} F_1(t) = \int_{-1}^1\! dx\; H(x,\eta,t)\,,\qquad
F_2(t) = \int_{-1}^1\! dx\; E(x,\eta,t)\,.
\end{eqnarray}
The Dirac form factor is normalized by $F_1(t=0)=1$,
cf.~Eq.~(\ref{Fix-Nor}). The asymptotic drop off for large $-t$ is
estimated to be $1/(-t)^2$, according to field theoretically
inspired \cite{DreYan69} and phenomenological \cite{Wes70} model
counting rules and the perturbative analysis \cite{BroFar75}. This
suggest to set $p=1$. We use  the charge radius squared $R^2= 6
dF_1(t)/dt|_{t=0}$ and the anomalous magnetic moment
$\kappa=F_2(t=0)$ of the proton to pin down the remaining two
parameters. The following plausible mass values yield an agreement
with experimental measurements, cf.~Ref.~\cite{Sch92}:
\begin{eqnarray}
p=1\,, \quad\lambda = 0.75\,{\rm GeV}\,, \quad m=0.45\,{\rm
GeV}\,\quad \Rightarrow\quad R=0.76\, {\rm fm} \,,\quad
\kappa=1.78\,. \label{Par-Spe}
\end{eqnarray}
\begin{figure}[t]
\begin{center}
\mbox{
\begin{picture}(250,105)(0,0)
\put(-123,60){\rotatebox{90}{$F_1(t)$}} \put(70,113){(a)}
\put(-110,0){\insertfig{7.5}{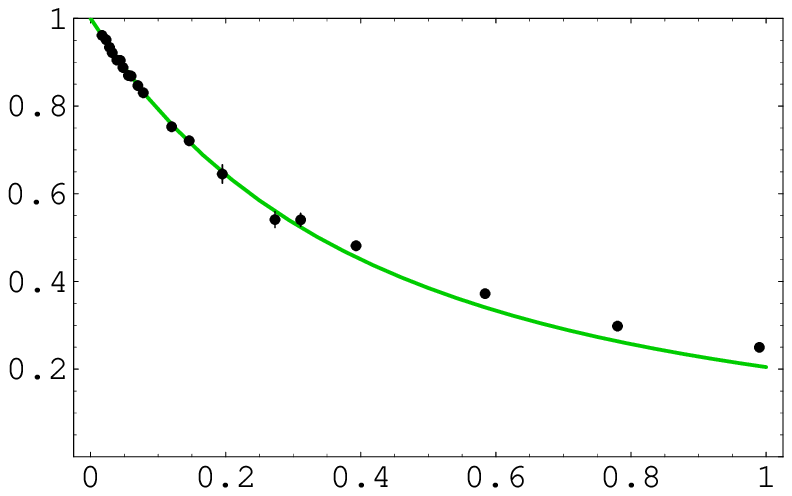}} \put(60,-10){$-t\, [{\rm
GeV}^2]$} \put(320,113){(b)}
\put(130,30){\rotatebox{90}{$(-t)F_2(t)/\kappa F_1(t)$}}
\put(150,0){\insertfig{7.3}{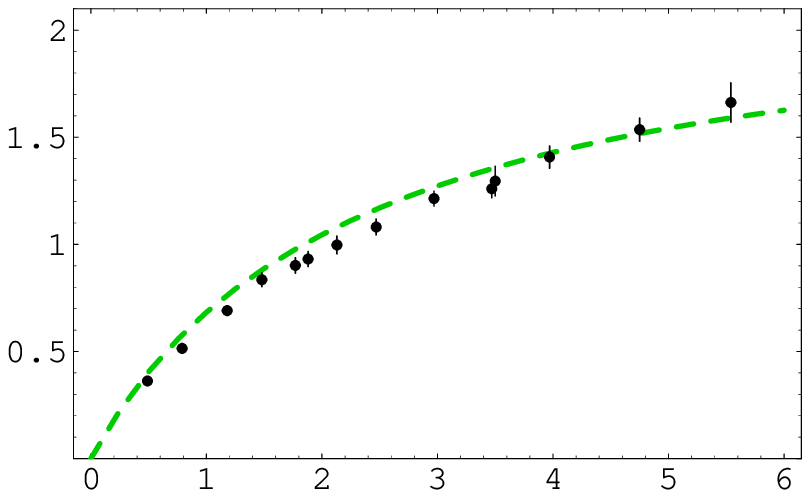}} \put(320,-10){$-t\, [{\rm
GeV}^2]$}
\end{picture}
}
\end{center}
\vspace{-2mm} \caption{ \label{Fig-FF} \small The form factor
$F_1(t)$ (a) and the ratio $ (-t) F_2(t)/\kappa F_1(t)$ with a
cut-off for the endpoint singularities (b). Data are taken from
Refs.~\cite{Hohetal76,Gayetal01,Punetal05}.}
\end{figure}
The form factor $F_1(t)$ fairly agrees with experimental data as
shown in Fig.\ \ref{Fig-FF}(a) and behaves for $-t \lessapprox
100\, {\rm GeV}^2$ as $1/(-t)^2$, however, in the large $-t$
asymptotics it drops faster.  We would conclude, in agreement with
aforementioned counting rules, that $F_2(t)$ decreases  as
$1/(-t)^{3}$. This follows from counting the powers of ${\bf
\Delta}_\perp$ in Eq.\ (\ref{Def-Ep}), arising from ${\bf
k}_\perp$-dependence of wave functions
(\ref{Def-LC-WF1}-\ref{Def-LC-WF3}), and the vanishing of the
overlap integral  for $E$  if  spherical symmetry is restored in
the asymptotic ${\bf \Delta}^2_\perp \to \infty$. However, we
observe that the limit ${\bf \Delta}^2_\perp \to \infty$ can not
be taken before the ${\bf k}_\perp$ integration, since divergences
appear and render the integral to be infinite. This behavior shows
up also in the DD (\ref{Def-SpeFun-e}), where the limit $-t \to
\infty$ causes end-point singularities at $z=\pm (1-y)$.  Hence,
we effectively find a $F_2(t) \sim 1/(-t)^{2}$ behavior and thus
naive power counting  fails. We remind that the assumptions for
the counting rules were carefully spelled out
\cite{DreYan69,Wes70,BroFar75}.

End-point singularities arise also in the perturbative evaluation
of $F_2$  and their regularization yields a logarithmical
modification of the $1/(-t)^3$ scaling
\cite{BelJiYua03,BroHwaKar03}. The experimental measurements
indicate a  $F_2(t)/F_1(t) \propto 1/\sqrt{-t}$ scaling
\cite{Jonetal99,Gayetal01,Punetal05}, which might be also
interpreted as a logarithmical modified $F_2(t)/F_1(t) \propto
1/(-t)$ scaling. Interestingly, the anomalous break down of the
naive power counting in our scalar diquark model leads to a
$F_2(t)/F_1(t) \propto {\rm const.}$ scaling. Applying the recipe
of end-point regularization to our result, i.e., imposing the
constraint $0.33 \le 1-y-|z|$, we easily can fit the experimental
data as demonstrated in Fig.~\ref{Fig-FF}(b). This exercise should
not be considered as a serious attempt to explain experimental
data, rather as a demonstration showing that fitting can be done
easily. One might wonder whatever this failure of the scalar
diquark model is related to the oversimplified spin coupling and
could be cured by inclusion of an axial-vector diquark, or it
might simply reflect a wrong implementation for the large ${\bf
k}^2_\perp$ behavior of the LC wave functions.

Our model also predicts TMDs and their ${\bf
k}_{\perp}$-integrated PDFs outcome. Note that the TMD concept has
numerous fundamental issues, see, e.g., Ref.~\cite{ColQui07}. We
might define here a unpolarized valence-like TMD in terms of the
LC wave functions (\ref{Def-LC-WF1}-\ref{Def-LC-WF3}) overlap,
where $\zeta=0$ and $t=0$:
\begin{eqnarray}
\label{Def-IntParDen} q(x, {\bf k}_{\perp})= {g^2 M^{4 p}
\left[(xM+m)^2+{\bf k}_{\perp}^2\right]\ (1-x)^{2p+1}\over
\left[{\bf k}_{\perp}^2 +(1-x)m^2+x{\lambda}^2 -x(1-x)M^2
\right]^{2p+2}}\,.
\end{eqnarray}
At large ${\bf k}^2_{\perp}$ they fall off with $1/({\bf
k}_{\perp}^2)^{2p+1}$ and are suppressed in the large $x$ region
by $(1-x)^{2p+1}$. Our model is similar in spirit to the spectator
model utilized in Ref.~\cite{JakMulRod97} for the evaluation of
${\bf k}_\perp$-(un)integrated parton densities and fragmentation
functions. There the fermionic propagators are replaced by
quark-diquark form factors with a cut-off mass $\Lambda$, while in
our case they are take to be on-shell. Replacing $m$ by $\Lambda$,
we find that Eq.~(\ref{Def-IntParDen}) is identical with the
result (80) in Ref.~\cite{JakMulRod97}.

The valence-like PDF is obtained from (\ref{Def-IntParDen}) and is
related to the GPD $H$:
\begin{eqnarray}
q(x) \equiv H(x,\eta=0,t=0) = \frac{\pi g^2 M^{4p+2}}{2 p(2p+1)}
\frac{\frac{(2 p-x+1) m^2}{M^2}+\frac{4 p x m}{M}+\frac{x \lambda
^2}{M^2}+x (2 p x+x-1) }{ \left(x \lambda ^2+(1-x) \left(m^2-M^2
x\right)\right)^{2 p+1} } (1-x)^{2 p+1}  . \label{Def-ParDen}
\end{eqnarray}
For $p=1$ we obtain the generic behavior of parton densities at
large $x$. Note that the form factor $F_1(t)$ falls at large $-t$
with $(-t)^{-p-1}$. Hence, setting $p=n-1$ we confirm the known
counting rules. The momentum fraction carried by the  valence
quark combination $u-d$,
\begin{eqnarray} \langle x \rangle =
\int_0^1\! dx\, x\, q(x)\,,\; q\equiv q_{u_{\rm val}}-q_{d_{\rm
val}}\,,
\end{eqnarray}
is $\langle x \rangle \approx 0.26$ with our parameter
specification (\ref{Par-Spe}). It nearly agrees with the value in
the Gl\"uck, Reya, and Vogt parameterization \cite{GluReyVog98},
given at a low input scale $\mu_0^2 = 0.4\, {\rm GeV}^2$ (to
perturbative next-to-leading order accuracy): $\langle x
\rangle_{u_{\rm val}} - \langle x \rangle_{d_{\rm val}} \approx
0.24$. This amazing coincidence of momentum fractions should be
considered rather as an accident.
For $x$ going to zero, the PDF (\ref{Def-ParDen}) approaches a
constant. This behavior we consider as an unrealistic feature of
the diquark model. The small $x$-region, i.e., the high-energy
limit might be understood in the Regge picture as an exchange of mesons
in the $t-$channel which leads to the expectation that the
valence-like parton densities behave as $x^{-\alpha(0)}$, where
$\alpha(0)$ is the (effective) intercept of the meson trajectory.
From the $s$-channel view, which we take, the true small
$x$-behavior arises by summing up all Fock state components.

We come now to the GPDs. First, we comment on our duality
assumption between the partonic $s$-channel and mesonic like
$t$-channel exchange. We could have added to the central region a
term
\begin{eqnarray}
D(x/\eta,t) =  \theta(|x| \le |\eta|)\,
d\!\left(\frac{x}{\eta},t\right),\quad\mbox{with}\quad
d\!\left(1,t\right)=d\!\left(-1,t\right)=0
\end{eqnarray}
that entirely lives in the central region, vanishes at the cross
over point and is anti-symmetric in $x$. An explicit GPD
evaluation from the parton number changing processes confirms that
such a term is absent and so the underlying field theoretical
model respects duality.   However, such a $D$-term is needed to
cure the common DD representation for the GPD $E$ or $H$
\cite{PolWei99}:
\begin{eqnarray}
\label{Def-SpeRep1} \left\{{H\atop E}\right\} (x,\eta,t) =
\int_{0}^1\!dy\; \int_{-1+y}^{1-y}\! dz\;
 \delta(x-y-\eta z )\;  \left\{{h\atop e}\right\}^{\rm Rad}(y,z,t) \pm  D(x/\eta,t).
\end{eqnarray}
Our finings suggest that both $e^{\rm Rad}(y,z,t)$ and
$d(x/\eta,t)$  have a cross talk. Indeed, they are two different
projections of our DD (\ref{Def-SpeFun-e});  the $D$-term is
extracted from $\eta\to\infty$ with fixed $x/\eta$
\cite{BelMueKirSch00}:
\begin{eqnarray}
\label{Cal-Dterm} d(x,t) = x \int_{0}^{1-|x|}\! dy \;   \frac{N
\left(\frac{m}{M} + y\right) ((1-y)^2-x^2)^p}{\left[(1-y)
\frac{m^2}{M^2}+ y \frac{\lambda^2}{M^2} - y(1-y) -((1-y)^2-x^2)
\frac{t}{4 M^2}\right]^{2p+1}}\,.
\end{eqnarray}
Interestingly, the coefficients in its Gegenbauer expansion
\cite{GoePolVan01} are rather small and sign alternating:
\begin{eqnarray}
\label{Cal-Dterm1} d(x,t=0) \approx (1-x^2)\left[ 0.345\,
C_1^{3/2}(x) -0.163\, C_3^{3/2}(x) + 0.026\, C_5^{3/2}(x) +\cdots
\right].
\end{eqnarray}
The first coefficient is in astonishing agreement with the
estimate of the chiral soliton model \cite{Wak07}, given as 0.33
at a intrinsic scale of $\mu_0^2 \sim 0.36\, {\rm GeV}^2$.  Note
that the flavor singlet quark $D$-term in the chiral soliton model
is predicted to be negative and large in magnitude
\cite{GoePolVan01,SchBofRad02,Wak07}.

In confronting  GPDs with experimental data, one usually
factorizes\footnote{Such an ansatz might be improved by replacing
the effective Regge intercept in the quark densities by the
$t-$dependent Regge trajectory, which will only partly remove the
factorized $t$-dependence.\label{foonot3}} the $t$-dependence in
the DD ansatz \cite{Rad00a,VanGuiGui98,GoePolVan01} (VGG refers to
the popular code of Vanderhaeghen, Guichon, and Guidal):
\begin{eqnarray}
\label{Def-SpeRep1ans} h^{\rm VGG}(y,z,t) =  F_1(t)
\frac{q(y)}{1-y} \Pi^{\rm VGG}(z/(1-y))\,, \quad \Pi^{\rm VGG}(z)
= \frac{ \Gamma \left(b+\frac{3}{2}\right)}{ \sqrt{\pi } \Gamma
(b+1)} (1-z^2)^b.
\end{eqnarray}
The even profile function $\Pi^{\rm VGG}(z)$ is  chosen to be
convex and normalized to $\int_{-1}^1\!dz\, \Pi^{\rm VGG}(z)=1$.

\begin{figure}[t]
\begin{center}
\mbox{
\begin{picture}(250,120)
\put(-120,50){\rotatebox{90}{$H(x,\eta,t)$}}
\put(-90,0){\insertfig{7}{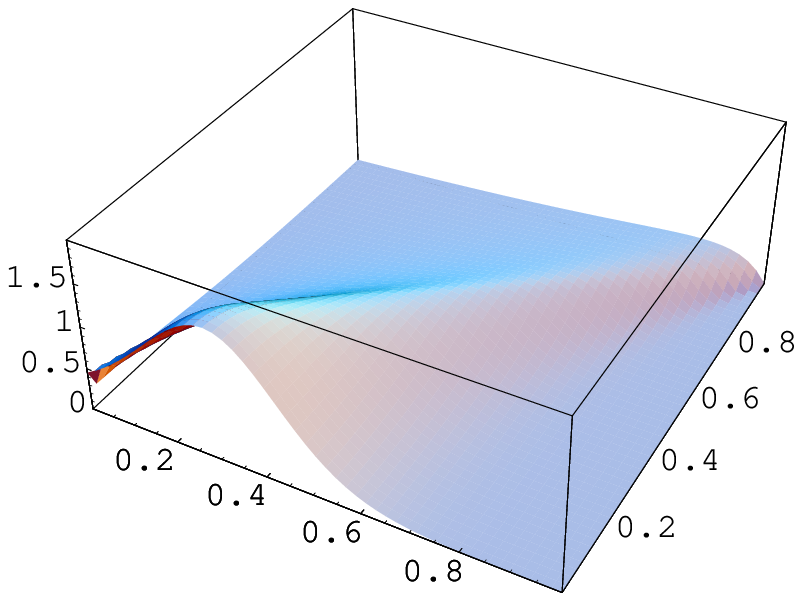}} \put(-30,10){$x$}
\put(0,130){(a)} \put(95,35){$\eta$}
\put(130,0){\insertfig{7}{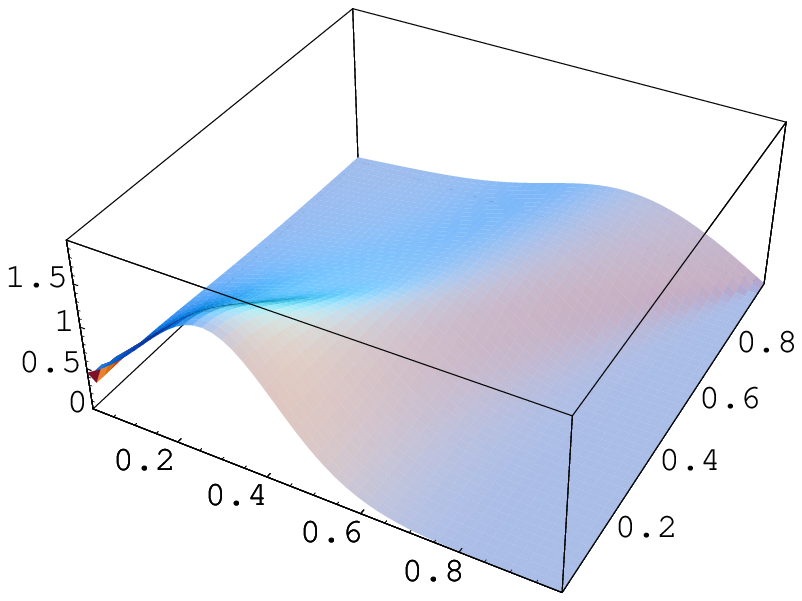}} \put(190,10){$x$}
\put(315,35){$\eta$} \put(220,130){(b)}
\end{picture}
}
\end{center}
\vspace{-8mm} \caption{ \label{Fig-ShaGPD} \small The GPD
$H(x,\eta,t)$ in the spectator model
(\ref{Def-SpeRep},\ref{Def-SpeFun-e},\ref{Def-SpeFun-eh}) (a) and
factorized VGG ansatz (\ref{Def-SpeRep1},\ref{Def-SpeRep1ans}) (b)
with the parameter values in (\ref{Par-Spe}). The momentum
transfer squared is set to  $t=t_{\min}-0.25\,{\rm GeV}^2$.}
\end{figure}
\begin{figure}[t]
\begin{center}
\mbox{
\begin{picture}(250,260)(0,0)
\put(-115,170){\rotatebox{90}{$H(\eta,\eta,t)/H(\eta,\eta,0)$}}
\put(-100,150){\insertfig{7.5}{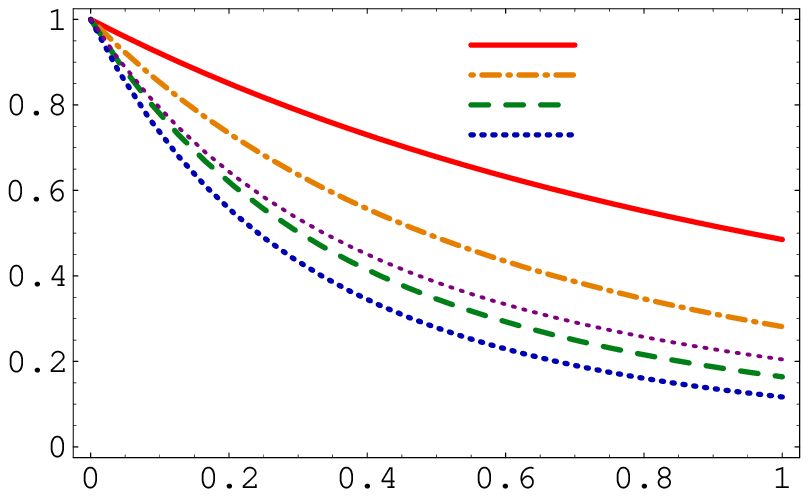}} \put(100,140){$-t$}
\put(58,269){\tiny $\eta=0.7$} \put(58,261){\tiny
$\phantom{\eta}=0.5$} \put(58,253){\tiny $\phantom{\eta}=0.3$}
\put(58,245){\tiny $\phantom{\eta}=0.1$} \put(90,260){(a)}
\put(125,175){\rotatebox{90}{$\displaystyle \frac{\int_0^1\!dx\;
x^n\; H(x,0,t)}{\int_0^1\!dx\; x^n\; H(x,0,0)}$}}
\put(160,150){\insertfig{7.5}{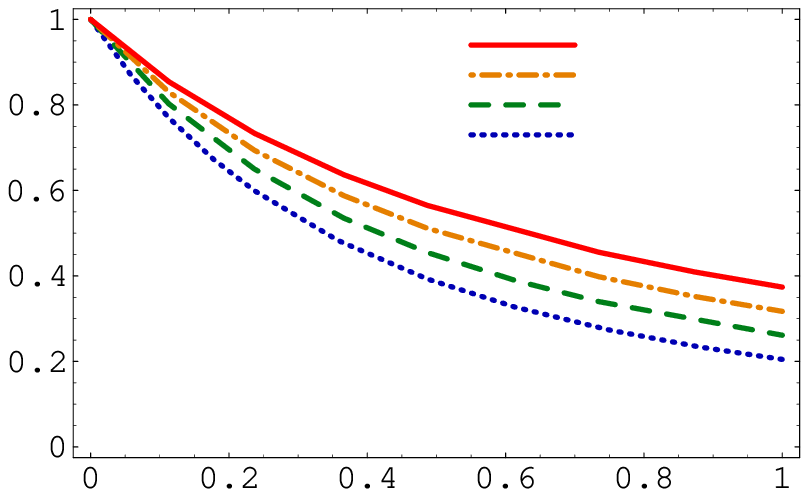}} \put(318,269){\tiny
$n=3$} \put(318,261){\tiny $\phantom{n}=2$} \put(318,253){\tiny
$\phantom{n}=1$} \put(318,245){\tiny $\phantom{\eta}=0$}
\put(350,260){(b)} \put(360,140){$-t$}
\put(-115,50){\rotatebox{90}{$ \Im{\rm m}\, {\cal H}(\xi,t) $}}
\put(-100,0){\insertfig{7.5}{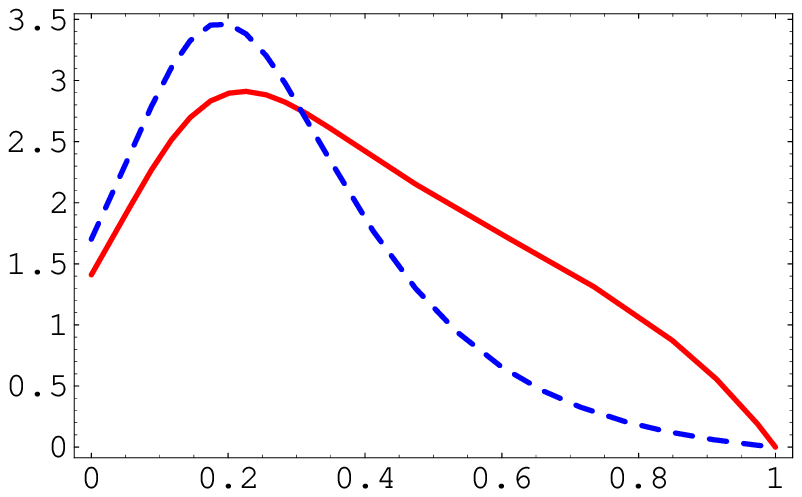}} \put(100,-10){$\xi$}
\put(90,100){(c)} \put(140,50){\rotatebox{90}{$ \Re{\rm e}\, {\cal
H}(\xi,t) $}} \put(160,0){\insertfig{7.5}{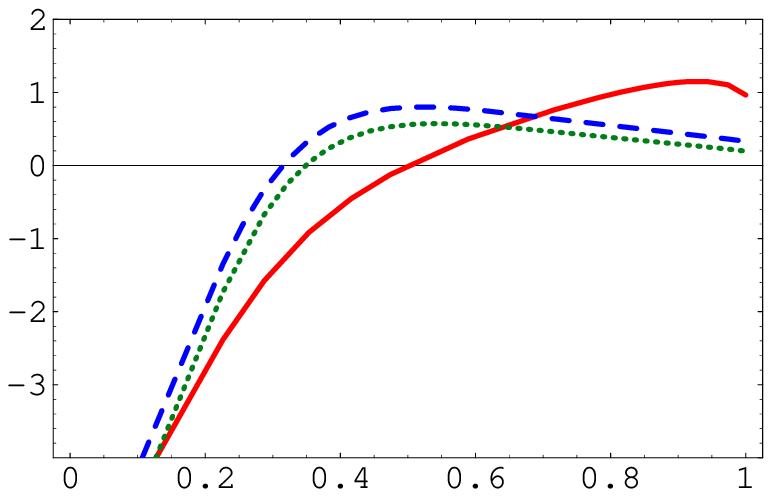}}
\put(360,-10){$\xi$} \put(200,100){(d)}
\end{picture}
}
\end{center}
\vspace{-2mm} \caption{\label{Fig-GPD-t-dep}\small  The
$t$-dependence of $H(x,\eta,t)/H(x,\eta,t=0)$  at $x=\eta$ in the
$t$-factorized ansatz (\ref{Def-SpeRep1},\ref{Def-SpeRep1ans})
(thin dotted) and the spectator model within $\eta= \{0.1 ({\rm
dotted}), 0.3 ({\rm dashed}), 0.5 ({\rm dashdotted}), 0.7 ({\rm
solid})\}$ (a) and for the   Mellin moments $n=\{0 ({\rm
dotted}),1 ({\rm dashed}),2 ({\rm dashdotted}) ,3 ({\rm solid})\}$
at $\eta=0$  in the spectator model (b). Imaginary (c) and real
(d) part of the amplitude (\ref{Def-CalH}) versus $\xi$, arising
from the spectator model (solid) and factorized ansatz (dashed),
where the dotted curve shows the real part without $D$-term.}
\end{figure}

Let us compare the spectator model
(\ref{Def-SpeRep},\ref{Def-SpeFun-e},\ref{Def-SpeFun-eh}) with the
$t$-factorized VGG ansatz
(\ref{Def-SpeRep1},\ref{Def-SpeRep1ans}). The latter we make from
the PDF (\ref{Def-ParDen}) and $D$-term (\ref{Cal-Dterm1}),
multiplied with the form factor (\ref{Def-FF}). For the profile
function we take $\Pi^{\rm VGG}(z)=(3/4)(1-z^2)$, which also
appears in our DD (\ref{Def-SpeFun-eh}) with $p=1$ at $t=0$,
cf.~Ref.~\cite{Rad00a}, and employ in both models the parameters
(\ref{Par-Spe}). In Fig.~\ref{Fig-ShaGPD} we show the shape of
GPDs versus $x$ and $\eta$, where the momentum transfer squared is
set to $t=t_{\rm min}(\xi)-0.25$. The differences between the
spectator model (a) and the $t$-factorized (b) GPDs are clearly
visible at larger values of $\eta$, where the later ones are
broader in their $x$ distribution and have a smaller value [see
also below Fig.~\ref{Fig-GPD-t-dep}(c)] on the trajectory
$x=\eta$, compared to the former ones. The important difference
between the two models is in their $t$-dependence. In
Fig.~(\ref{Fig-GPD-t-dep}) we display in panel (a) the
$t$-dependence of the ratio $H(\eta,\eta,t)/H(\eta,\eta,0)$ on the
trajectory $x=\eta$ for various values of $\eta$. Obviously, the
$t$-dependence is flattering out for larger values of $\eta$,
while in the small $\eta$ region it even becomes steeper, compared
to the $t$-factorized GPD ansatz (dashed line). Analogously, we
find for the  ratio $\int_0^1\!dx\, x^n
H(x,0,t)/\int_0^1\!dx\, x^n H(x,0,0)$ of Mellin moments at $\eta=0$ that the
$t$-dependence becomes flatter with increasing spin $n$ as
shown in panel (b). Such a behavior was also seen in lattice
calculations, see Ref.~\cite{Hagetal07} and references therein.

As in the case of PDFs, the small $x$-behavior of GPDs in a
spectator model should be considered as unrealistic. Since at
$x=\eta$ the momentum fraction $X'$ vanishes in the overlap
representation (\ref{Def-Hp},\ref{Def-Ep}), we realize that on
this trajectory realistic GPDs can be obtained only if the small
$X$ behavior of the wave functions is understood. This also means
that one has to sum up all Fock state components, see discussion
in Refs.~\cite{DieFelJakKro98,JiMisRad06}. Note that already the
evolution to leading order in the flavor singlet sector knows
about the small $x$ behavior, however, in the non-singlet sector
one has to perform a resummation of $t$-channel exchanges,
perhaps, along the line as suggested in Ref.~\cite{ErmGreTro00}.
For larger values of $x$ on $x=\eta$ one is forced to understand
at the same time the large and small $X$ behavior of the wave
functions and their interference.

Having this warning in mind, we have now a look at the resulting
amplitude, which appears in the hard exclusive photon or
$\rho^{0}$ electroproduction to leading order of perturbative QCD:
\begin{eqnarray}
\label{Def-CalH} {\cal H}(\xi,t) =\int_{-1}^1\!dx\;
\left[\frac{1}{\xi-x -i \epsilon} - \frac{1}{\xi+x -i
\epsilon}\right] H(x,\eta=\xi,t)\,.
\end{eqnarray}
The imaginary and real part of the amplitude versus $\xi$ are
shown in Fig.~\ref{Fig-GPD-t-dep} (c) and (d), respectively, where
we set $t=t_{\rm min}-0.25$. Note that the imaginary part is given
by $\pi H(x=\xi,\xi,t)$ and provides us the GPD shape on the
trajectory $x=\xi$. The differences between the spectator (solid
curve) and $t$-factorized (dashed curve) models are obvious. Both
the imaginary and real part for the spectator model is (much) more
enhanced at large $\xi$. For $\xi\to 1$ the imaginary part
(\ref{Fig-GPD-t-dep}(c), solid) behaves as $(1-\xi)$, while for
the $t$-factorized GPD ansatz an additional suppression factor
appear: $(1-\xi) (-t)^{-2} \lesssim (1-\xi) (-t_{\rm min})^{-2}
\sim (1-\xi)^3$. On the other hand, we observe that at smaller
values of $\xi$ the imaginary part of the spectator model is
getting smaller. We stress, however, that small $\xi$-physics is
not implemented in the model. The sign change of the real part in
panel (d), somewhere in the valence quark region, is a feature
that is also observed in other valence-like GPD ansatze. The
position of the zero, however, is floating and strongly depends on
the chosen ansatz. At smaller values of $\xi$ the real parts of
both models approach to each other. The dotted curve shows the
$t$-factorized ansatz without $D$-term, leading only to a slight
constant shift.

Concerning the results of our spectator model, we conclude that
the $t$-dependence of cross sections for deeply electroproduction
of mesons in the large $x_{\rm Bj}$ region is getting flatter and
that the cross section could be  (much) larger than the prediction
from a $t$-factorized GPD ansatz, cf.~Figs.~\ref{Fig-GPD-t-dep}
(c) and (d). This is  consistent with preliminary measurements  of
the $e^-p(P_1) \to e^- p(P_2) \rho^{(0)}$ cross section, where the
$t$-slope decreases if  larger values of $x_{\rm Bj}$ and $Q^2$
are approached \cite{GuiMor07pricom}.  We find, for instance, that
the slope for the amplitude square $|{\cal H}(\xi,t)|^2$,
parameterized as $e^{b t}$, decreases from $b\approx 3.5\, {\rm
GeV}^{-2}$ at $x_{\rm Bj}=0.2$ and $Q^2=2\, {\rm GeV}^2$ to
$b\approx 1.5\, {\rm GeV}^{-2}$ at $x_{\rm Bj}=0.6$ and $Q^2=5\,
{\rm GeV}^2$.

\vspace{2mm}

\noindent iv.~In this paper we derived from the overlap
representation of LC wave functions the valence-like GPDs and
their relatives: proton form factors, TMDs, and PDFs. We
generalized the LC wave functions from a scalar diquark spectator
model in such a way that the non-manifest Lorentz behavior of the
LC wave functions is respected. This is the key to obtain the DDs
from the overlap representation in the partonic $s$-channel and
then to restore the full GPD support.

Our model fairly describes the Dirac form factor $F_1(t)$ and the
proton  anomalous magnetic moment comes out correctly by a natural
choice for the constituent quark and diquark masses. However, the
model fails in the description of the $t$-dependence for the Pauli
form factor $F_2(t)$. Namely, its naive power counting for the
large $t$-behavior is spoiled by end-point singularities that
appears at the branch point $-t=\infty$. Unpolarized valence-like
TMDs and PDFs, also obtained in Ref.~\cite{JakMulRod97}, have the
expected generic behavior at large momentum fraction $x$ and the
average value $\langle x \rangle$ fairly agrees with the result of
Ref. \cite{GluReyVog98}, given at a low input scale.

The GPD models satisfy by construction the positivity and
polynomiality constraints. The found DD representations
(\ref{Def-SpeRep}) naturally completes the polynomiality condition
and avoids so a `misleading' $D$-term phenomenology. Another
important characteristic property of the model is that the
$t$-dependence of GPDs, resulting from the DDs
(\ref{Def-SpeFun-eh}) and (\ref{Def-SpeFun-e}), is washed out at
large $x$. This behavior is rooted in the fact that the variables
$t$ and $x$ are intrinsically correlated because the transverse
and longitudinal degrees of freedom in the wave functions are tied
by hidden Lorentz symmetry. This flattering of $t$-dependence also
appears on the trajectory $x=\eta$ and therefore it can be
confronted with experimental measurements.  Note that the
$t$-behavior  of GPDs is only  tested in lattice calculations for
$\eta=0$. From the overlap representation it is clear that the
$t$-dependence of GPDs and the ${\bf k}_\perp$-dependence of TMDs
are closely related to each other, since both arise from the ${\bf
k}_\perp$-dependence of wave functions, a recent discussion is
given in Ref.~\cite{MeiMetGoe07}.

It is in the nature of a spectator model that it fails to describe
the small momentum fraction behavior, which arises from the
summation over all partonic Fock state components. Hence, there is
a potential problem for the predicting power of such models for
GPDs on the trajectory $x=\eta$, accessible in experiments. Here a
GPD arises from the interference of the LC wave function at the
extreme limit of vanishing momentum fraction with that of a
non-vanishing momentum fraction, controlled by the skewness
parameter. Therefore, even the limit $x=\eta\to 1$ is rather
intricate \cite{Yua03}.

It remains a challenging task to construct GPDs that satisfy all
theoretical constraints and are flexible enough for a `global' fit
of experimental data. For the time being, we might suggest to
adopt the $t$-dependence in the ansatz for the DD  and improve its
failure at small $y$, i.e, small $x$ for the resulting GPD, by
hand. Guided by Eq.~(\ref{Def-SpeFun-e}), we would suggest for
$e$, e.g., the ansatz:
\begin{eqnarray}
\label{Def-SpeFun-e-Imp} e\!\left(y,z,t\right) =
\frac{q_E(y,t)}{(1-y)}  \frac{
  N(b,p,\alpha) \left[(1-y) \frac{m^2}{M^2}+ y \frac{\lambda^2}{M^2}
-y(1-y)\right]^{P}}{\left[(1-y) \frac{m^2}{M^2}+ y
\frac{\lambda^2}{M^2} - y(1-y) -((1-y)^2-z^2) \frac{t}{4
M^2}\right]^{P}}  \frac{[(1-y)^2-z^2]^{b}}{(1-y)^{2b+1}} \,.
\label{zz2}
\end{eqnarray}
The Regge improved PDF analog, generically parameterized by its
large and small $x$ behavior
\begin{eqnarray}
q_E(x,t)= \kappa \frac{(1-\alpha(0)) \Gamma(2-\alpha(t)+\beta) }{
\Gamma(2-\alpha(t))\Gamma(1+\beta)}  x^{-\alpha(t)} (1-x)^\beta\,,
\quad \int_0^1\!dx\, q_E(x,t=0)= \kappa\,,
\end{eqnarray}
where $q_E(x,t)$ is normalized at $t=0$ to the anomalous magnetic
moment $\kappa$. The normalization
\begin{eqnarray}
N(b,P,\alpha)= \frac{ \Gamma \left(b+\frac{3}{2}\right) \Gamma(2
-\alpha (0)+\beta) \Gamma (2-2 P -\alpha (t)+\beta) }{ \sqrt{\pi }
\Gamma (b+1) \Gamma (2-2 P -\alpha(0)+\beta) \Gamma(2-\alpha
(t)+\beta)}
\end{eqnarray}
is introduced in such a way that the $t$-dependence in the form
factor {\em roughly} factorizes as  $1/(1-\alpha(t))$, resulting
from the Regge behavior, times an impact form factor, behaving as
$(-t)^{-P}$ for $t\to -\infty$ . The parameter $b$ adjusts the
power behavior of $E(\xi,\xi,t) \sim (1-\xi)^{1+b}$ at large $\xi$
and fixed $t$. The parameters should be fixed from fitting of
observables;  their generic values read in accordance with Regge
phenomenology, counting rules
\cite{DreYan69,Wes70,BroFar75,Yua03}, and the spectator model:
$$
\alpha(t)\sim 0.5 + 0.9\, t\, {\rm GeV}^{-2},\;\; \beta\sim 5,\;\;
P \sim 2,\;\; b \sim 1,\;\; \lambda \sim 0.8\, {\rm GeV},\;\;
m\sim 0.4\, {\rm GeV}\,.
$$
The model features, we spelled out here in momentum fraction
representation, are also implemented in the Mellin space GPD
ansatz \cite{MueSch05,KumMuePas07}. We emphasize that the ansatz
(\ref{zz2}) is still not optimal, since a flexible adjustment of
the resulting magnitude for the amplitude at given $t$, in
particular at small $\xi$, is not incorporated and positivity
constraints are no more automatically satisfied. A more detailed
discussion of improved GPD ansatze should be given somewhere else.

\vspace{5 mm}

\noindent D.S.H.~thanks the Institute for Theoretical Physics II
at the Ruhr-University Bochum, in particular K.~Goeke, for the
hospitality at the stages of this work. Both authors enjoyed many
stimulating discussions with the members of the group. We are
grateful to P.~Pobylitsa, who inspired us to our studies,
and to M.~Guidal for illuminating discussions about experimental
data.    This work was supported in part by the BMBF (Federal
Ministry for Education and Research), contract FKZ 06 B0 103,  by
the International Cooperation Program of the KICOS (Korea
Foundation for International Cooperation of Science \&
Technology), and by the European Union  project ``Joint Research
Activity five: {\em Generalized Parton Densities}''.

\end{document}